\documentclass{jetpl}
\twocolumn

\usepackage{graphicx}
\usepackage{cite}
\usepackage{amsmath}
\usepackage{amsfonts}
\lat

\title{Magnetic flux detection with an Andreev Quantum Dot}

\rtitle{Magnetic flux detection with an Andreev Quantum Dot}

\sodtitle{Magnetic flux detection with an Andreev Quantum Dot}

\author{I.A.~Sadovskyy$^*$\/\thanks{sadovsky@itp.ac.ru},
        G.B.~Lesovik$^*$,
        G.~Blatter$^\nabla$}

\rauthor{I.A.~Sadovskyy, G.B.~Lesovik, G.~Blatter}

\sodauthor{Sadovskyy, Lesovik, Blatter}

\address{
  $^*$L.D.~Landau Institute for Theoretical Physics, Russian Academy of Sciences,
  119334, Kosygina st., 2, Moscow, Russia\\
  $^\nabla$Theoretische Physik, Schafmattstrasse 32,
  ETH-Zurich, CH-8093 Z\"urich, Switzerland
}

\dates{25 May 2007}{*}

\begin{document}

\abstract{The charge of the subgap states in an Andreev quantum
dot (AQD; this is a quantum dot inserted into a superconducting
loop) is very sensitive to the magnetic flux threading the loop.
We study the sensitivity of this device as a function of its
parameters for the limit of a large superconducting gap $\Delta$.
In our analysis, we account for the effects of a weak Coulomb
interaction within the dot. We discuss the suitability of this
setup as a device detecting weak magnetic fields.}

%%% PACS numbers
%http://publish.aps.org/PACS/

\PACS{73.21.La,  % Quantum dots
      74.45.+c,  % Proximity effects; Andreev effect; SN and SNS junctions
      07.55.Ge}  % Magnetometers for magnetic field measurements

\maketitle

% ------------ Introduction ----------------------------

\paragraph{Introduction.}
The Josephson effect~\cite{Josephson} has been intensively studied
during the past 45 years; its main characteristic is the presence
of a tunable non-dissipative current when two bulk superconductors
are joined via a normal or insulating layer and subjected to a
superconducting phase difference $\varphi$.  Recently, it has been
realized that in a metallic junction the charge of the normal
island in between the superconducting leads depends on the
superconducting phase difference $\varphi$ as
well~\cite{Sad,Engstrom}. This dependence is sufficiently
strong~\cite{Sad} to use this effect in a magnetic flux detector,
although our estimates below give a sensitivity somewhat below the
sensitivity of the best SQUIDs.

Usually, small magnetic fields are measured by superconducting
quantum interference devices
(SQUIDs)~\cite{Kleiner,Clarke_Braginski}. While SQUIDs are based
on the dependence of the Josephson current on the superconducting
phase difference $\varphi$ (and hence on the magnetic flux $\Phi$
threading the loop), here we propose to use the charge-dependence
in an Andreev quantum dot for the flux measurement.  As shown in
Ref.~\cite{Sad}, the charge $Q$ of a single-channel Andreev
quantum dot can be fractional $-|e|<Q<|e|$ and depends on
$\varphi$ (here $e=-|e|$ is the charge of one electron).

The charge of an Andreev quantum dot can be measured by a
sensitive charge detector, e.g., by a single-electron transistor
(SET). Today, the best single electron transistors have a
sensitivity of the order of $10^{-5} \, |e|/\sqrt{\mathrm{Hz}}$
(e.g., see~\cite{aassime_01}). Using results of Ref.~\cite{Sad},
simple estimates tell that an AQD can convert a change in flux
$\delta \Phi$ to a change in charge $\delta Q$ with a ratio
$\delta Q / \delta \Phi \sim 2 |e|/\Phi_0$, where $\Phi_0 = 2\pi
\hbar / 2|e|$ is the superconducting flux. Assuming a
superconducting loop area $\sim 1$~mm$^2$, we obtain the
sensitivity $10^{-14} \, \mathrm{T}/ \sqrt{ \mathrm{Hz}}$, which
is comparable with the sensitivity $10^{-14} \div 10^{-15} \,
\mathrm{T}/\sqrt{\mathrm{Hz}}$ of today's best
SQUIDs~\cite{Kleiner,Clarke_Braginski}. Below, we study in detail
the sensitivity ratio $\delta Q / \delta \Phi$.

% ------------ Setup ----------------------------

\paragraph{Setup.}
\begin{figure}[b]
  \centerline{\includegraphics[width=6.5cm]{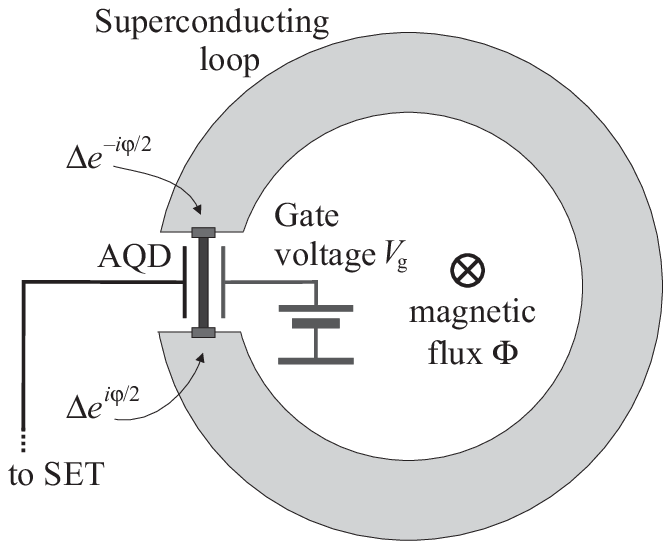}}
  \caption[]{Fig.~\ref{fig:loop}. Andreev quantum dot
  inserted into the superconducting loop. The Andreev quantum dot
  is connected to a single electron transistor (SET) and a gate
  electrode through capacitive couplings. The flux $\Phi$
  produces a phase difference $\varphi = 2\pi \Phi/\Phi_0$
  across the Andreev quantum dot. The charge of the AQD can be
  tuned by the gate voltage $V_\mathrm{g}$ and the flux $\Phi$
  threading the loop.}
  \label{fig:loop}
\end{figure}
Our Andreev quantum dot is realized by a small metallic dot
connecting two superconducting banks joined in a loop, see
Fig.~\ref{fig:loop}. Our AQD is assumed to be a quasi
one-dimensional normal metal (N) island separated from the
superconductors (S) by thin insulator layers (I), generating
normal scattering on top of the Andreev scattering characteristic
of the normal-metal superconductor junction. The position of the
normal resonance in this SINIS system can be tuned by the gate
voltage $V_\mathrm{g}$ applied to the normal region of the AQD.
The magnetic flux threading the loop $\Phi$ induces a
superconducting phase drop $\varphi$ across the AQD. Since the
phase drop in the bulk superconductor is negligible as compared to
the phase drop $\varphi$ across the AQD one may relate the latter
to the flux $\Phi$ threading the loop, $\varphi = 2\pi
\Phi/\Phi_0$. In order to measure the charge trapped on the AQD, a
single electron transistor is capacitively coupled to the normal
metal island. Experimentally, such AQDs have recently been
fabricated by coupling carbon nanotubes to superconducting
banks~\cite{Kouwenhoven,Kouwenhoven_2,Bouchiat,cnt_K}. In the
following, we concentrate on the properties of the key element in
the setup --- the Andreev quantum dot.

% ------------ Energy and charge of AQD without interaction ----------------------------

\paragraph{Energy and charge of the AQD without Coulomb interaction}
The Andreev states give rise to new opportunities for tunable
Josephson devices, e.g., the Josephson transistor
\cite{JT,Shumeiko,Kuhn}; here, we are interested in their charge
properties. We will consider the case of one transverse channel
such that the problem effectively becomes one dimensional. We
consider the case of a large separation $\delta_{\rm
\scriptscriptstyle N}$ between the resonances in the associated
NININ problem (where the superconductors S have been replaced by
normal metal leads N), $\delta_{\rm\scriptscriptstyle N} \gg
\Delta$, such that a single Andreev level
$\varepsilon_{\rm\scriptscriptstyle A}$ is trapped within the gap
region. We are interested in sufficiently well isolated dots with
a small width $\Gamma_{\rm\scriptscriptstyle N}$ of the associated
NININ resonance, $\Gamma_{\rm\scriptscriptstyle N} \ll \Delta$. In
this section, we neglect charging effects
$E_{\rm\scriptscriptstyle C} = 0$. In summary, our device operates
with energy scales $\Gamma_{\rm\scriptscriptstyle N} \ll \Delta
\ll \delta_{\rm\scriptscriptstyle N}$.

The resonances in the NININ setup derive from the eigenvalue
problem $\mathcal{\hat H}_0 \Psi=E\Psi$ with $\mathcal{\hat H}_0 =
-\hbar^2\partial_x^2/2m + U(x) - \varepsilon_{\rm
\scriptscriptstyle F}$ with the potential $U(x) =
U_\mathrm{ps,\,1}(x+L/2) + U_\mathrm{ps,\,2}(x-L/2)] +
eV_\mathrm{g} \theta(L/2-|x|)]$ describing two
point-scatterers\footnote{The Heaviside function $\theta(x)=1$ for
$x>0$ and $\theta(x)=0$ for $x \leqslant 0$.} (with transmission
and reflection amplitudes $T_l^{1/2}e^{\chi^t_l}$,
$R_l^{1/2}e^{\chi^r_l}$; $R_l=1-T_l$, $l=1,$ $2$) and the effect
of the gate potential $V_\mathrm{g}$, which we assume to be small
as compared to the particle's energy $E$ (measured from the band
bottom in the leads), $eV_\mathrm{g} \ll E$. Resonances then
appear at energies $E_n = \varepsilon_L (n\pi - \chi^r_1/2 -
\chi^r_2/2)^2$; they are separated by $\delta_{\rm
\scriptscriptstyle N}^n =(E_{n+1}-E_{n-1})/2 \approx 2E_n/n$ and
are characterized by the width $\Gamma_{\rm \scriptscriptstyle
N}^n = T \delta_{\rm \scriptscriptstyle N}^n / \pi\sqrt{R}$, where
$\varepsilon_L=\hbar^2/2mL^2$. The bias $V_\mathrm{g}$ shifts the
resonances by $eV_\mathrm{g}$; we denote the position of the
$n$-th resonance relative to $\varepsilon_{\rm \scriptscriptstyle
F}$ by $\varepsilon_{\rm \scriptscriptstyle N}^n = E_n +
eV_\mathrm{g} - \varepsilon_{\rm \scriptscriptstyle F}$. In the
following, we choose a specific resonance in the gap by selecting
an appropriate $n$ and drop the index $n$, $\varepsilon_{\rm
\scriptscriptstyle N}^n \to \varepsilon_{\rm \scriptscriptstyle
N}$, $\delta_{\rm \scriptscriptstyle N}^n \to \delta_{\rm
\scriptscriptstyle N}$, $\Gamma_{\rm \scriptscriptstyle N}^n \to
\Gamma_{\rm \scriptscriptstyle N}$.

We go from a normal- to an Andreev dot by replacing the normal
leads with superconducting ones. In order to include Andreev
scattering in the SINIS setup, we have to solve the Bogoliubov-de
Gennes equations (we choose states with $\varepsilon_{\rm
\scriptscriptstyle A} \geqslant 0$)
\begin{eqnarray}
\left[
  \begin{array}{cc}
  \mathcal{\hat H}_0 (x)  & {\hat \Delta} (x) \\
  {\hat \Delta}^* (x)     & -\mathcal{\hat H}_0 (x)
  \end{array}
\right]
  \!
\left[
  \begin{array}{c}
  \!\!u \!\!\! \\ \!\! v \!\!\!
  \end{array}
\right]
  = \varepsilon_{\rm \scriptscriptstyle A}
  \left[
  \begin{array}{c}
  \!\!u \!\!\! \\ \!\! v \!\!\!
  \end{array}
\right],
  \label{BdG}
\end{eqnarray}
with the pairing potential ${\hat \Delta}(x) = \Delta[\theta(-x-L/2)
e^{-i\varphi/2} + \theta(x-L/2) e^{i\varphi/2}]$; $u(x)$ and $v(x)$ are the
electron- and hole-like components of the wave function. The discrete states
trapped below the gap derive from the quantization condition (in Andreev
approximation)
\begin{multline}
  (R_1 + R_2) \cos\!\Big( 2\pi \frac{
    \varepsilon_{\rm\scriptscriptstyle A}
  }{
    \delta_{\rm\scriptscriptstyle N}
  }\Big) -
  4\sqrt{R_1R_2} \, \sin^2 \alpha \, \cos\!\Big( 2\pi \frac{
    \varepsilon_{\rm\scriptscriptstyle N}
  }{
    \delta{\rm\scriptscriptstyle N}
  }\Big) +
\\
  +T_1 T_2 \cos\varphi
  = \cos\!\Big(
    2\alpha - 2\pi \frac{
      \varepsilon_{\rm\scriptscriptstyle A}
    }{
      \delta_{\rm\scriptscriptstyle N}
    }
  \Big) +
  R_1 R_2 \cos\!\Big(
    2\alpha + 2\pi\frac{
      \varepsilon_{\rm\scriptscriptstyle A}
    }{
      \delta_{\rm\scriptscriptstyle N}
    }
  \Big).
  \nonumber
\end{multline}
The phase $\alpha = \arccos(\varepsilon_{\rm\scriptscriptstyle
A}/\Delta)$ is acquired at an ideal NS boundary due to Andreev
reflection with $\varphi=0$; the above formula can be directly
obtained using results from Refs.~\cite{Kuhn,Chtchelkatchev}.

We concentrate on the regime $\Gamma_{\rm\scriptscriptstyle N},
|\varepsilon_{\rm\scriptscriptstyle N}| \ll \Delta$, the so-called
$\Delta \to \infty$ limit. In this limit, the quantization
condition can be expanded and we obtain the expression ($A$ is the
asymmetry parameter)
\begin{equation}
  \varepsilon_{\rm\scriptscriptstyle A} =
  \sqrt{ \varepsilon_{\rm\scriptscriptstyle N}^2 +
  \varepsilon_{\rm\scriptscriptstyle \Gamma}^2 },
  \label{Andreev_en}
\end{equation}
where
\begin{equation}
  \varepsilon_{\rm\scriptscriptstyle \Gamma} =
  \frac{\Gamma_{\rm\scriptscriptstyle N}}{2}
  \sqrt{ \cos^2 \frac{\varphi}{2} + A^2 }, \;\;\;
  A = \frac{|T_1-T_2|}{2\sqrt{T_1T_2}}.
  \label{E_Gamma}
\end{equation}
The energy $\varepsilon_{\rm\scriptscriptstyle A}$ of the Andreev
state is phase sensitive when $\varepsilon_{\rm\scriptscriptstyle
N}$ is close to the chemical potential,
$|\varepsilon_{\rm\scriptscriptstyle N}| \lesssim
\Gamma_{\rm\scriptscriptstyle N}$, which can be achieved by tuning
the gate potential $V_\mathrm{g}$. In the limit $\Delta \to
\infty$, both the $u(x)$ and $v(x)$ components of the wave
function are nonzero only in the normal region,
\begin{equation}
  \left[
  \begin{array}{c}
    \!\!u(x) \!\!\! \\ \!\! v(x) \!\!\!
  \end{array}
  \right]
= \left\{
  \begin{array}{cl}
    0, & \!|x| > L/2, \\
    \left[
    \begin{array}{c}
        \!\! C^{\rightarrow}_\mathrm{e} e^{ik_\mathrm{e}x} +
        C^{\leftarrow}_\mathrm{e} e^{-ik_\mathrm{e}x} \!\!\!
      \\
        \!\! C^{\leftarrow}_\mathrm{h} e^{ik_\mathrm{h}x} +
        C^{\rightarrow}_\mathrm{h} e^{-ik_\mathrm{h}x} \!\!\!
    \end{array}
    \right] \!, &
    \! |x| < L/2,
  \end{array}
  \right.
 \nonumber
\end{equation}
where $k_{\mathrm{e}, \mathrm{h}} = [2m (
\varepsilon_{\rm\scriptscriptstyle F} \pm
\varepsilon_{\rm\scriptscriptstyle A} )]^{1/2}/\hbar$ are the wave
vectors of electrons and holes, respectively. The coefficients are
defined by $C^{\rightarrow}_\mathrm{e,h} =
C^{\leftarrow}_\mathrm{e,h} = [( 1 \pm
\varepsilon_{\rm\scriptscriptstyle N} /
\varepsilon_{\rm\scriptscriptstyle A} )/4L]^{1/2}$.

The ground state of the system is the state $|0\rangle$ with
energy
\begin{equation}
  \varepsilon_0 =
  \varepsilon_{\rm\scriptscriptstyle N} -
  \varepsilon_{\rm\scriptscriptstyle A}
  \label{bU_0}
\end{equation}
(counted from the Fermi energy $\varepsilon_{\rm
\scriptscriptstyle F}$), where we have subtracted the energy of
filled resonances below the Fermi surface; the latter are not
modified by the superconductivity in the leads and hence do not
depend on the phase $\varphi$. The first excited state with one
Bogoliubov quasiparticle is doubly degenerate in spin
$|1_\uparrow\rangle = {\hat a}_\uparrow^\dag |0\rangle$,
$|1_\downarrow\rangle = {\hat a}_\downarrow^\dag |0\rangle$ and
has energy $\varepsilon_1 = \varepsilon_0 +
\varepsilon_{\rm\scriptscriptstyle A} =
\varepsilon_{\rm\scriptscriptstyle N}$. The doubly excited state
with two quasiparticles $|2\rangle = {\hat a}_\uparrow^\dag {\hat
a}_\downarrow^\dag |0\rangle$ has an energy $\varepsilon_2 =
\varepsilon_0 + 2\varepsilon_{\rm\scriptscriptstyle A} =
\varepsilon_{\rm\scriptscriptstyle N} +
\varepsilon_{\rm\scriptscriptstyle A}$.

The charge of the state $|\nu\rangle$ ($\nu=0$, $1_\uparrow$,
$1_\downarrow$, $2$) can be obtained by differentiation of the
corresponding energy $\varepsilon_\nu$ with respect to the gate
voltage, $q_\nu = \partial \varepsilon_\nu / \partial
V_\mathrm{g}$, or by averaging the charge operator ${\hat Q} = e
\sum_\sigma \int_{-L/2}^{L/2} {\hat\Psi}_{\sigma}^\dag(x)
{\hat\Psi}_{\sigma}^{\phantom \dag}(x) \, dx$ over the state
$|\nu\rangle$, $q_\nu = \langle \nu | {\hat Q} | \nu \rangle$.
Both methods give the identical results
\begin{multline}
  q_0 = e \Big(1 - \frac{
    \varepsilon_{\rm\scriptscriptstyle N}
  }{
    \varepsilon_{\rm\scriptscriptstyle A}
  } \Big), \;
  q_1 = e, \;
  q_2 = e \Big(1 + \frac{
    \varepsilon_{\rm\scriptscriptstyle N}
  }{
    \varepsilon_{\rm\scriptscriptstyle A}
  } \Big).
  \label{q_123}
\end{multline}
Below, we will also need the off-diagonal matrix elements of the
charge operator ${\hat Q}$; the only non-vanishing term is $q_{02}
= \langle 0 | {\hat Q} | 2 \rangle = e(1- \varepsilon_{\rm
\scriptscriptstyle N}^2 / \varepsilon_{\rm\scriptscriptstyle A}^2
)^{1/2}$.

% ------------ AQD with Coulomb interaction ----------------------------

\paragraph{AQD with Coulomb interaction}
\begin{figure}[b]
   \centerline{\includegraphics[width=7.9cm]{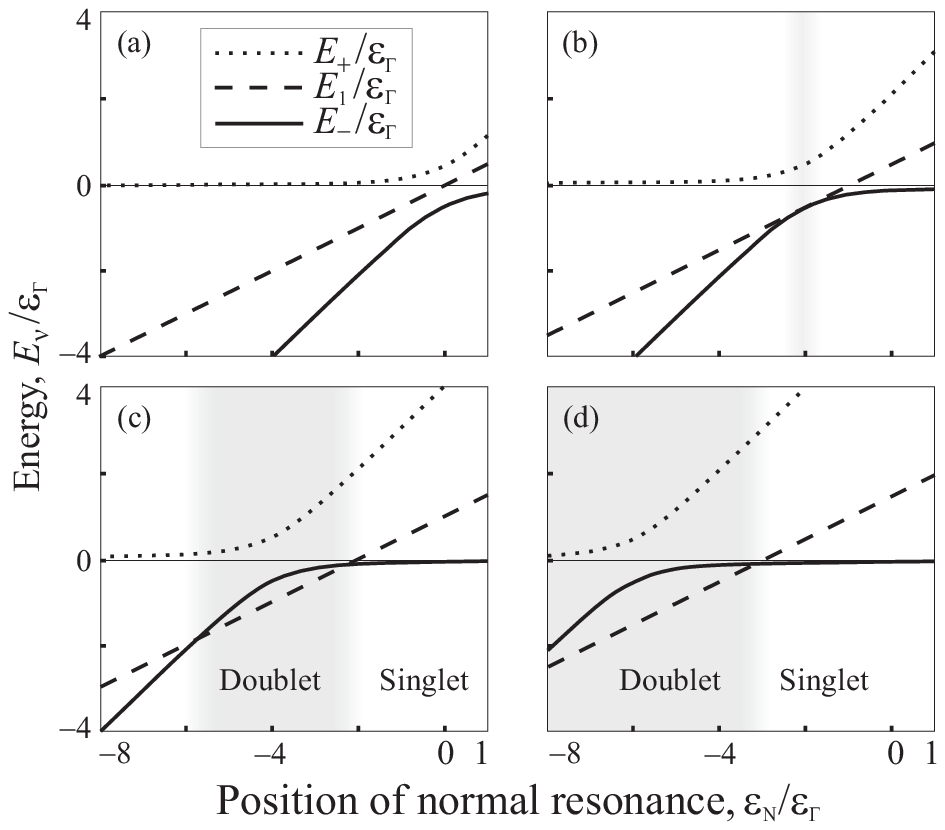}}
   \caption[]{Fig.~\ref{fig:energies}.
      Energies $E_-$ (solid line), $E_1$ (dashed line), and $E_+$ (dotted
      line) are plotted versus the position of normal resonance. All energies
      are given in units of $\varepsilon_{\rm\scriptscriptstyle \Gamma}$, cf.\
      \eqref{E_Gamma}. The Coulomb energy is $E_{\rm\scriptscriptstyle C}=0$
      for (a), $E_{\rm\scriptscriptstyle C}=\varepsilon_{\rm\scriptscriptstyle
      \Gamma}$ for (b), $E_{\rm\scriptscriptstyle C}=2\varepsilon_{\rm
      \scriptscriptstyle \Gamma}$ for (c), and $E_{\rm\scriptscriptstyle
      C}=3\varepsilon_{\rm \scriptscriptstyle \Gamma}$ for (d).  In accordance
      with formula~\eqref{Kr_d} the doublet region appears when $E_{\rm
      \scriptscriptstyle C} > \varepsilon_{\rm\scriptscriptstyle \Gamma}$, see
      (b--d).  In the filled region the ground state of the system is a
      doublet; the width of this region is $2 (E_{\rm \scriptscriptstyle C}^2
      - \varepsilon_{\rm \scriptscriptstyle \Gamma}^2)^{1/2}$, the edges of
      this region are spread due to the finite temperature $\Theta$.}
   \label{fig:energies}
\end{figure}
In order to find the effect of weak Coulomb interaction $E_{\rm
\scriptscriptstyle C} \ll \Delta$ in the limit $\Gamma_{\rm
\scriptscriptstyle N},$ $|\varepsilon_{\rm\scriptscriptstyle N}|
\ll \Delta$, we can disregard the continuous states with energies
above the superconducting gap $\Delta$ and assume that the four
levels of the discrete spectrum form the entire basis of the
system's Hilbert space\footnote{In realistic nanodevices the
Coulomb energy can be larger then $\Gamma_{\rm\scriptscriptstyle
N}$ and smaller or of the order of $\delta_{\rm\scriptscriptstyle
N}$, but in principle can be made much smaller than both
$\delta_{\rm\scriptscriptstyle N}$ and $\Delta$ (see the
discussion in~\cite{Kouwenhoven,Sad}).}. The interaction is given
by the operator
\begin{equation}
  {\hat V} = E_{\rm\scriptscriptstyle C} \frac{{\hat Q}^2}{e^2}.
\end{equation}
Given the basis with these four states, we can diagonalize the Hamiltonian
exactly. The non-zero matrix elements of the operator ${\hat V}$ are
\begin{multline}
  V_{00} = E_{\rm\scriptscriptstyle C} (q_0^2 + q_{02}^2)/e^2, \;
  V_{11} = E_{\rm\scriptscriptstyle C}, \; \\
  V_{22} = E_{\rm\scriptscriptstyle C} (q_2^2 + q_{02}^2)/e^2, \;
  V_{02} = 2 E_{\rm\scriptscriptstyle C} q_{02}/e.
  \label{V_ij}
\end{multline}
The energy levels are defined by the eigenvalue problem
\begin{eqnarray}
\left[ \begin{array}{cccc}
  \!\! {\tilde \varepsilon}_0 - E \!\!\! &  &  & \!\! V_{02} \!\!\!   \\
  & \!\! {\tilde \varepsilon}_{1\uparrow} - E \!\!\! &  &  \\
  &  & \!\! {\tilde \varepsilon}_{1\downarrow} - E \!\!\! &  \\
  \!\! V_{20} \!\!\!  &  &  & \!\! {\tilde \varepsilon}_2 - E \!\!\! \\
\end{array} \right]
  \!
\left[ \begin{array}{l}
  \!\! D_0 \!\!\!\! \\  \!\! D_{1\uparrow} \!\!\!\! \\ \!\! D_{1\downarrow}
  \!\!\!\! \\ \!\! D_2  \!\!\!\!
\end{array} \right]
  =0,
  \nonumber
\end{eqnarray}
where ${\tilde \varepsilon}_\nu = \varepsilon_\nu + V_{\nu\nu}$,
$\nu=0$, $1_\uparrow$, $1_\downarrow$, $2$. The energy of the
level with one Bogoliubov quasiparticle $|1\rangle$ is given by
the (shifted) constant
\begin{equation}
  E_1 = \varepsilon_{\rm\scriptscriptstyle N} + E_{\rm\scriptscriptstyle C},
  \label{E_1}
\end{equation}
and does not mix with the other states; furthermore, the spin
degeneracy of this Kramers doublet remains. The ground state
$|0\rangle$ and the doubly excited state $|2\rangle$ mix due to
Coulomb interaction and produce two new states, the singlet states
$|-\rangle$ and $|+\rangle$; $|\pm\rangle = D_0^\pm|0\rangle +
D_2^\pm|2\rangle$, $D_0^\pm/D_2^\pm = -V_{02}/({\tilde
\varepsilon}_0 - E_\pm)$, $|D_0^\pm|^2 + |D_2^\pm|^2 = 1$. The
energies of these new states are
\begin{equation}
  E_\pm = \varepsilon_{\rm\scriptscriptstyle N} + 2E_{\rm\scriptscriptstyle C} \pm
  \sqrt{(\varepsilon_{\rm\scriptscriptstyle N} + 2E_{\rm\scriptscriptstyle C})^2
  + \varepsilon_{\rm\scriptscriptstyle \Gamma}^2}.
  \label{E_pm}
\end{equation}

The energies of the doublet and singlet states depend on
$\varepsilon_{\rm \scriptscriptstyle N}$ and $\varphi$ in a
different way and may cross; thus the ground state can be formed
by either the singlet $|-\rangle$ or by the doublet $|1\rangle$.
The state $|+\rangle$ always remains the second excited state, see
Fig.~\ref{fig:energies}. When $E_{\rm\scriptscriptstyle C} >
\varepsilon_{\rm\scriptscriptstyle \Gamma} \geqslant \Gamma_{\rm
\scriptscriptstyle N} A/2 \equiv E_{\rm\scriptscriptstyle C}^*$
(with $A$ the asymmetry parameter) the ground state is the doublet
$|1\rangle$ in the region
\begin{equation}
  -2E_{\rm\scriptscriptstyle C} - \sqrt{E_{\rm\scriptscriptstyle C}^2
   - \varepsilon_{\rm\scriptscriptstyle \Gamma}^2}
  < \varepsilon_{\rm\scriptscriptstyle N} <
  -2E_{\rm\scriptscriptstyle C} + \sqrt{E_{\rm\scriptscriptstyle C}^2
   - \varepsilon_{\rm\scriptscriptstyle \Gamma}^2}
  \label{Kr_d}
\end{equation}
and remains $|-\rangle$ at all other values of
$\varepsilon_{\rm\scriptscriptstyle N}$~\cite{rozhkov_arovas}.

The origin of this level crossing can be traced to the different shifts in
energies with $E_{\rm\scriptscriptstyle C}$: While $E_1$ is shifted up by
$E_{\rm\scriptscriptstyle C}$, $E_-$ quickly approaches 0 with increasing
$E_{\rm\scriptscriptstyle C}$.  Note that the terms $\propto q_{02}^2$ and
$\propto q_{02}$ in the matrix elements $V_{\nu\nu'}$ lead to the crossing of
the energies $E_-$ and $E_1$, while preventing the crossing of the level $E_+$
with the others.

At the edge of the region~\eqref{Kr_d} a sharp singlet to doublet
crossover takes place, with a jump appearing as a function of
$\varepsilon_{\rm\scriptscriptstyle N}(V_\mathrm{g})$ or
$\varepsilon_{\rm\scriptscriptstyle \Gamma}(\varphi)$ in the
charge of the Andreev dot and in the current across (see below).
\begin{figure}[t]
  \centerline{\includegraphics[width=7.9cm]{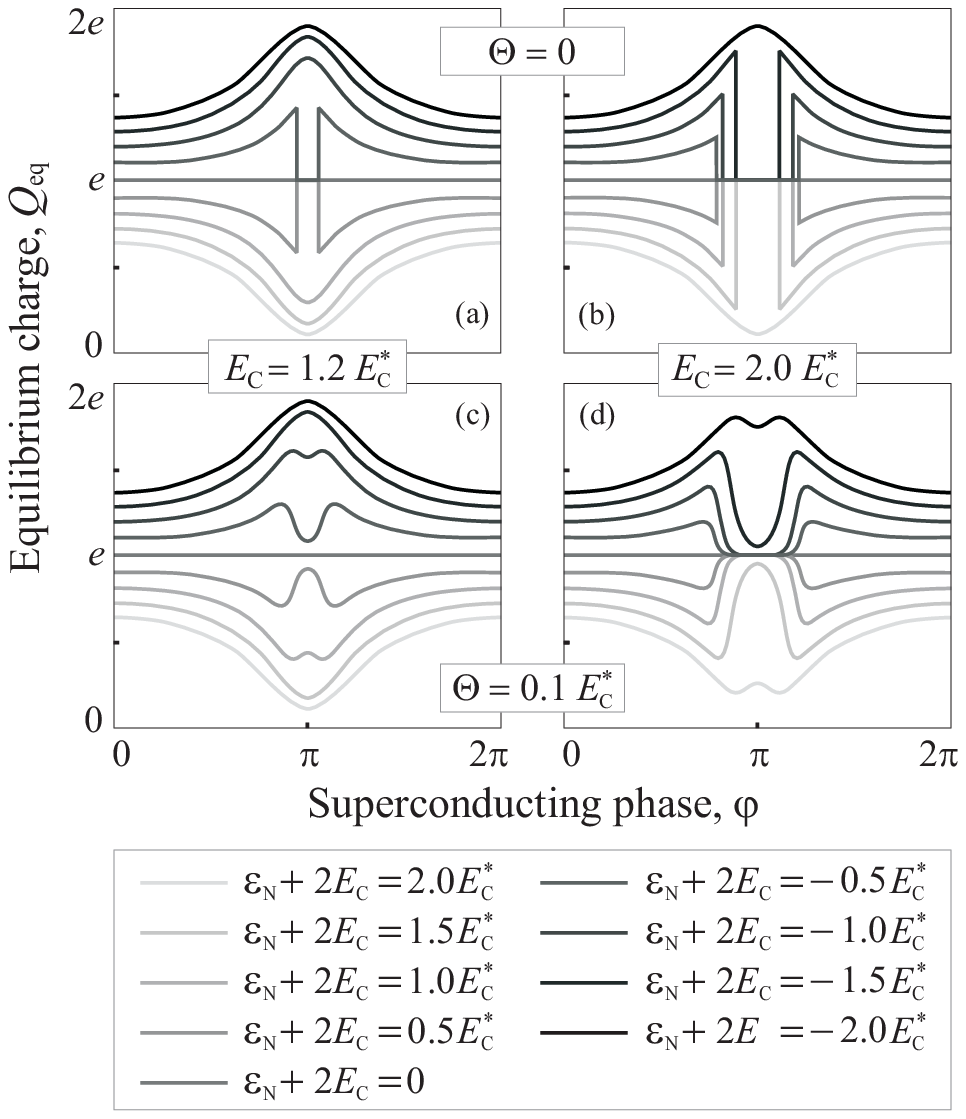}}
  \caption[]{Fig.~\ref{fig:charge}.
    Equilibrium charge $Q_\mathrm{eq}$~\eqref{Q_eq} versus superconducting
    phase difference $\varphi$. In (a) and (b) the temperature is zero
    (i.e.,  $Q_\mathrm{eq}$ represents ground state charge), in (c) and (d)
    the temperature is $\Theta = 0.1 E_{\rm\scriptscriptstyle C}^*$, where
    $E_{\rm\scriptscriptstyle C}^* \equiv \Gamma_{\rm\scriptscriptstyle N}
    A/2$.  The Coulomb energy is $E_{\rm\scriptscriptstyle C} = 1.2
    E_{\rm\scriptscriptstyle C}^*$ for (a) and (c), $E_{\rm\scriptscriptstyle
    C} = 2.0 E_{\rm\scriptscriptstyle C}^*$ for (b) and (d).  The asymmetry
    level of the dot is $A=0.2$. The features in the center of the plots
    corresponds to the Kramers doublet region~\eqref{Kr_d}. In (c) and (d) the
    border of the doublet region is smoothed by the temperature $\Theta$.}
   \label{fig:charge}
\end{figure}
The charges of the new states $|\mu\rangle$, ($\mu=1,$ $\pm$) can be
calculated as in the previous section, $Q_\mu = \partial E_\mu / \partial
V_{\mathrm g}$, and are given by
\begin{equation}
  Q_\pm = e \bigg(1 \pm \frac{\varepsilon_{\rm\scriptscriptstyle N}
  + 2E_{\rm\scriptscriptstyle C}}{\sqrt{(\varepsilon_{\rm\scriptscriptstyle N}
  + 2E_{\rm\scriptscriptstyle C})^2
  + \varepsilon_{\rm\scriptscriptstyle \Gamma}^2}} \bigg), \;\;\;
  Q_1 = e.
  \label{Q_m2p}
\end{equation}
The charge $Q_1$ is integer and does not fluctuate; the charges
$Q_\pm$ are fractional in the region
$|\varepsilon_{\rm\scriptscriptstyle N} +
2E_{\rm\scriptscriptstyle C}| \sim
\varepsilon_{\rm\scriptscriptstyle \Gamma}$ and fluctuate strongly
(see also the discussion of fluctuations in Ref.~\cite{Sad} where
Coulomb effects have been ignored)
\begin{multline}
  \delta Q_\pm \equiv
  [ \langle\pm {\hat Q}^2|\pm\rangle - \langle\pm {\hat Q}|\pm\rangle^2 ]^{1/2} =
  \\
  = e \frac{
    \varepsilon_{\rm\scriptscriptstyle \Gamma}
  }{
     \sqrt{(\varepsilon_{\rm\scriptscriptstyle N} + 2E_{\rm\scriptscriptstyle C})^2
     + \varepsilon_{\rm\scriptscriptstyle \Gamma}^2}
  }.
  \label{deltaQ}
\end{multline}
Note that the Coulomb interaction merely shifts the regime of
$\varepsilon_{\rm\scriptscriptstyle N}$ where the charges $Q_\pm$
are fractional. Everywhere outside the doublet region the ground
state charge is given by $Q_-$, while within the Kramers doublet
region the charge is pinned to the value $Q_1=e$. As illustrated
in Figs.~\ref{fig:charge}a and \ref{fig:charge}b, for
$E_{\rm\scriptscriptstyle C}
> E_{\rm\scriptscriptstyle C}^*$ a sharp crossover occurs and the charge
jumps by the value $\delta Q_\mathrm{cr} = Q_- - Q_1$.  This jump
is smeared at finite temperatures, see Figs.~\ref{fig:charge}c and
\ref{fig:charge}d. The groundstate charge is
\begin{equation}
  Q_\mathrm{gs} = e - e\frac{
    E_{\rm\scriptscriptstyle N}
  }{
    E_{\rm\scriptscriptstyle A}
  }\,
  \theta[
    E_{\rm\scriptscriptstyle C} < E_{\rm\scriptscriptstyle A}
  ],
  \nonumber
\end{equation}
where $E_{\rm\scriptscriptstyle A} =
[(\varepsilon_{\rm\scriptscriptstyle N} +
2E_{\rm\scriptscriptstyle C})^2 +
\varepsilon_{\rm\scriptscriptstyle \Gamma}^2]^{1/2}$ and
$E_{\rm\scriptscriptstyle N}=\varepsilon_{\rm \scriptscriptstyle
N} + 2E_{\rm\scriptscriptstyle C}$ denotes the energy of the
shifted normal state resonance. The equilibrium charge at finite
temperature $\Theta$ is
\begin{equation}
  Q_\mathrm{eq} = \frac{
    Q_- e^{-E_-/\Theta} + 2 Q_1 e^{-E_1/\Theta} + Q_+ e^{-E_+/\Theta}
  }{
    e^{-E_-/\Theta} + 2 e^{-E_1/\Theta} + e^{-E_+/\Theta}
  };
  \label{Q_eq}
\end{equation}
here and below we set Boltzmann's constant
$k_{\rm\scriptscriptstyle B} = 1$. The equilibrium charge as a
function of the superconducting phase $\varphi=2\pi \Phi/\Phi_0$
is shown in Fig.~\ref{fig:charge}.

The currents in the states $|\mu\rangle$ are defined by
relationship $J_\mu = \partial E_\mu / \partial \Phi$ which
provides the results
\begin{equation}
  J_\pm = \mp
  \frac{2\pi}{\Phi_0}
  \frac{
    \Gamma_{\rm\scriptscriptstyle N}^2 \sin\varphi
  }{
    16 E_{\rm\scriptscriptstyle A}
  }, \;\;\;
  J_1 = 0.
  \label{J_m2p}
\end{equation}
The groundstate current is
\begin{equation}
  J_\mathrm{gs} =
  \frac{2\pi}{\Phi_0}
  \frac{
    \Gamma_{\rm\scriptscriptstyle N}^2 \sin\varphi
  }{
    16 E_{\rm\scriptscriptstyle A}
  } \,
  \theta[
    E_{\rm\scriptscriptstyle C} < E_{\rm\scriptscriptstyle A}
  ];
  \nonumber
\end{equation}
note that the current vanishes throughout the doublet region. The
thermal equilibrium current is
\begin{equation}
  J_\mathrm{eq} =
  \frac{
    J_- e^{-E_-/\Theta} + J_+ e^{-E_+/\Theta}
  }{
    e^{-E_-/\Theta} + 2 e^{-E_1/\Theta} + e^{-E_+/\Theta}
  }.
  \label{J_eq}
\end{equation}

% ------------ Sensitivity ----------------------------

\paragraph{Differential sensitivity}
The differential sensitivity of the equilibrium charge to the
magnetic flux threading the superconducting loop is defined by the
absolute value of the derivative $\partial Q_\mathrm{eq} /
\partial\Phi$ taken at the given value of flux,\footnote{
    Note that the sensitivity of the charge-to-flux convertor $S \equiv
    S_{{\rm\scriptscriptstyle \Phi} \to {\rm\scriptscriptstyle Q}}$ coincides
    with the voltage-to-current sensitivity of the Josephson transistor
    described in Ref.~\cite{Kuhn} $S_{{\rm\scriptscriptstyle V} \to
    {\rm\scriptscriptstyle J}} = |\partial J_\mathrm{eq} / \partial V_\mathrm{g}|$.}
$S = |\partial Q_\mathrm{eq} / \partial \Phi|$. By
using~\eqref{Q_eq} we obtain
\begin{equation}
  S = \left| F_\Theta \, \frac{\partial Q}{\partial \Phi} +
  Q \, \frac{\partial F_\Theta}{\partial \Phi} \right|,
  \label{diff_sens}
\end{equation}
where $Q \equiv (Q_+ - Q_-)/2$, the derivative
\begin{equation}
  \frac{\partial Q}{\partial \Phi} =
  e \frac{2\pi}{\Phi_0} \, \frac{
    E_{\rm\scriptscriptstyle N}
    \Gamma_{\rm\scriptscriptstyle N}^2 \sin\varphi
  }{
    16 E_{\rm\scriptscriptstyle A}^3
  },
  \label{dQex}
\end{equation}
the function
\begin{multline}
  F_\Theta =
  \frac{
    e^{-E_+/\Theta} - e^{-E_-/\Theta}
  }{
    e^{-E_-/\Theta} + 2 e^{-E_1/\Theta} + e^{-E_+/\Theta}
  } =
  \\
  = -\frac{
    \sinh(E_{\rm\scriptscriptstyle A}/\Theta)
  }{
    \cosh(E_{\rm\scriptscriptstyle A}/\Theta) + e^{E_{\rm\scriptscriptstyle C}/\Theta}
  },
  \label{F}
\end{multline}
and its derivative
\begin{equation}
  \frac{\partial F_\Theta}{\partial \Phi}
  = \frac{
    e^{E_{\rm\scriptscriptstyle C}/\Theta} \sinh(E_{\rm\scriptscriptstyle A}/\Theta) + 1
  }{
    [\cosh(E_{\rm\scriptscriptstyle A}/\Theta) + e^{E_{\rm\scriptscriptstyle C}/\Theta}]^2
  } \,
  J_-.
  \label{F}
\end{equation}

As illustrated in Fig.~\ref{fig:charge} there are two intervals
where the $Q_\mathrm{eq}(\varphi)$ dependence is steep. As
$\varphi$ increases from $\varphi=0$, the charge increases
(decreases) and reaches a maximum (minimum). For
$E_{\rm\scriptscriptstyle C} < E_{\rm\scriptscriptstyle C}^*$ the
maximum (minimum) of the charge is always at $\varphi=\pi$, while
for $E_{\rm \scriptscriptstyle C} > E_{\rm\scriptscriptstyle C}^*$
the extremum splits and a second interval with a steep dependence
$Q_\mathrm{eq}(\varphi)$ emerges in between the two new extrema.
The first interval (interval I in what follows) corresponds to the
singlet state of the AQD, the second (interval II in what follows)
corresponds to the doublet state. We start with a description of
the first interval. We fix the parameters
$\Gamma_{\rm\scriptscriptstyle N}$, $A$, and
$E_{\rm\scriptscriptstyle C}$ and search for the maximum
sensitivity $S$ as a function of
$\varepsilon_{\rm\scriptscriptstyle N}$ and $\varphi$. The
non-trivial symmetries $Q_\mathrm{eq}(\varphi,\,
\varepsilon_{\rm\scriptscriptstyle N}) =
Q_\mathrm{eq}(2\pi-\varphi,\, \varepsilon_{\rm\scriptscriptstyle
N})$, $Q_\mathrm{eq}(\varphi,\, \varepsilon_{\rm\scriptscriptstyle
N}) - Q_\mathrm{eq}(\varphi,\, 0) = -Q_\mathrm{eq}(\varphi,\,
-\varepsilon_{\rm\scriptscriptstyle N} - 4E_{\rm\scriptscriptstyle
C}) + Q_\mathrm{eq}(\varphi,\, 0)$ allow us to restrict the search
to the region $0 \leqslant \varphi \leqslant \pi$,
$\varepsilon_{\rm\scriptscriptstyle N} + 2E_{\rm\scriptscriptstyle
C} \geqslant 0$. Subsequently, we analyze the maximum as a
function of $E_{\rm\scriptscriptstyle C}$ keeping $A$ and
$\Gamma_{\rm \scriptscriptstyle N}$ constant.

{\it Interval I:} For $E_{\rm\scriptscriptstyle C} <
[3(1+A^2)/(1+2A^2)]^{1/2} E_{\rm\scriptscriptstyle C}^*$ and zero
temperature $\Theta = 0$ the sensitivity is determined by the
derivative $\partial Q / \partial\Phi$~\eqref{dQex}. The function
$|\partial Q /
\partial\Phi|$ has a maximum at $\varepsilon_{\rm\scriptscriptstyle N} +
2E_{\rm\scriptscriptstyle C} = [(1+A^2)/(1+2A^2)]^{1/2}
E_{\rm\scriptscriptstyle C}^*$ and $\varphi = \pi -
2\arcsin[A/(1+2A^2)^{1/2}]$, where the differential sensitivity is
given by
\begin{equation}
  S_\mathrm{max}^\mathrm{I} = |e| \frac{2\pi}{\Phi_0}
  \frac{1}{6\sqrt{3} A \sqrt{1+A^2}}.
  \label{S0_max}
\end{equation}
One observes that the smaller $A$ is, the larger is the
sensitivity. In other words, a symmetric SINIS structure provides
a better sensitivity $S_\mathrm{max}^\mathrm{I}(A \to 0) \to
\infty$, but at the same time the region in $\varphi$ with this
large sensitivity vanishes as $A \to 0$. When $\Theta \ll
E_{\rm\scriptscriptstyle C}^*$ the sensitivity is nearly
independent of temperature.

In the opposite case $E_{\rm\scriptscriptstyle C} \geqslant
[3(1+A^2)/(1+2A^2)]^{1/2} E_{\rm\scriptscriptstyle C}^*$ the
doublet region covers all of the interval I and the maximum at
zero temperature is always realized at the edge of the doublet
region~\eqref{Kr_d}, with a sensitivity given by
\begin{multline}
  S_\mathrm{max}^\mathrm{I} =
  |e| \frac{2\pi}{\Phi_0} \frac{
    \Gamma_{\rm\scriptscriptstyle N}^3
  }{
    48 \sqrt{3} E_{\rm\scriptscriptstyle C}^3
  }
\times \\ \times
  \sqrt{2(\lambda^2-\lambda+1)^{3/2} - (\lambda+1)(\lambda-2)(2\lambda-1)}
  \label{S1_max}
\end{multline}
realized at $\varepsilon_{\rm\scriptscriptstyle N} +
2E_{\rm\scriptscriptstyle C} = (\Gamma_{\rm\scriptscriptstyle
N}/2) \{[2\lambda-1+(\lambda^2-\lambda+1)^{1/2}]/3\}^{1/2}$ and
$\varphi = 2\arccos\{[\lambda + 1 -
(\lambda^2-\lambda+1)^{1/2}]/3\}^{1/2}$, where $\lambda =
(E_{\rm\scriptscriptstyle C}^2 - {E_{\rm\scriptscriptstyle
C}^*}^2)/(\Gamma_{\rm\scriptscriptstyle N}/2)^2$. This result
reduces to
\begin{equation}
  S_\mathrm{max}^\mathrm{I} \approx
  |e| \frac{2\pi}{\Phi_0} \frac{
    \Gamma_{\rm\scriptscriptstyle N}^2
  }{
    16 E_{\rm\scriptscriptstyle C}^2
  }
\end{equation}
in the limit $E_{\rm\scriptscriptstyle C} \gg
\Gamma_{\rm\scriptscriptstyle N}$, and remains approximately
correct for $E_{\rm\scriptscriptstyle C} \approx
\Gamma_{\rm\scriptscriptstyle N}/2$. For $E_{\rm\scriptscriptstyle
C} \gg \Gamma_{\rm\scriptscriptstyle N}$, the maximum sensitivity
is reached at $\varepsilon_{\rm\scriptscriptstyle N} +
2E_{\rm\scriptscriptstyle C} \approx E_{\rm\scriptscriptstyle C} -
\Gamma_{\rm\scriptscriptstyle N}^2 / 16 E_{\rm\scriptscriptstyle
C}$ and $\varphi \approx \pi/2 + \Gamma_{\rm\scriptscriptstyle
N}^2 / 16 E_{\rm\scriptscriptstyle C}^2$.
\begin{figure}[t]
  \centerline{\includegraphics[width=6.8cm]{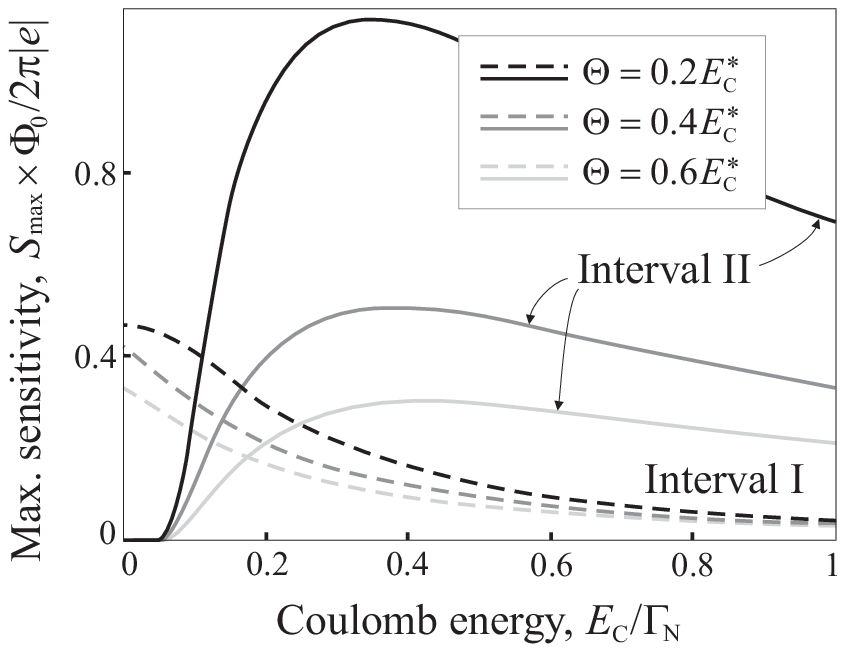}}
  \caption[]{
    Fig.~\ref{fig:max_sens}.
    Maximum of the differential sensitivity $S_\mathrm{max}$ (absolute value) in
    the interval I (dashed lines) and in the interval II (solid lines) versus
    Coulomb energy $E_{\rm\scriptscriptstyle C}$ at the asymmetry level $A=0.2$,
    $E_{\rm\scriptscriptstyle C}^* / \Gamma_{\rm\scriptscriptstyle N} = 0.1$.
    The temperature varies from $\Theta = 0.2 E_{\rm\scriptscriptstyle C}^*$ up
    to $\Theta = 0.6 E_{\rm\scriptscriptstyle C}^*$.
  }
  \label{fig:max_sens}
\end{figure}

{\it Interval II:} At zero temperature there is a jump in the
charge at the edges of interval II and thus the sensitivity
diverges in these points. A finite temperature smears the jump and
the sensitivity becomes finite. If $E_{\rm\scriptscriptstyle C}
\gg \Theta$, $\Gamma_{\rm\scriptscriptstyle N},$
$E_{\rm\scriptscriptstyle C}^*$, the sensitivity $S$ reaches the
maximum near the point $\varepsilon_{\rm\scriptscriptstyle N} +
2E_{\rm\scriptscriptstyle C} = E_{\rm\scriptscriptstyle C}$,
$\varphi = \pi/2$ where it equals to
\begin{equation}
  S_\mathrm{max}^\mathrm{II} \approx
  |e| \frac{2\pi}{\Phi_0} \frac{
    \Gamma_{\rm\scriptscriptstyle N}^2
  }{
    64 E_{\rm\scriptscriptstyle C} \Theta
  }.
\end{equation}
The expression for $S_\mathrm{max}^\mathrm{II}$ is too cumbersome
for an arbitrary Coulomb energy $E_{\rm\scriptscriptstyle C}$ and
we plot the numerical result
$S_\mathrm{max}^\mathrm{II}(E_{\rm\scriptscriptstyle C})$ in
Fig.~\ref{fig:max_sens}. In the same plot, we also present the
maxima of the sensitivity from the interval I. One easily notes
that for a large Coulomb interaction the charge jump smeared by
temperature provides the sharper $Q_\mathrm{eq}(\varphi)$
dependence.

% ------------ Conclusion ----------------------------

\paragraph{Conclusion.}
In this article, we have pointed out that the $\varphi$-dependence
of the charge trapped within an Andreev quantum dot may be used
for the implementation of a new type of magnetometer which
operates along the pathway `magnetic flux--AQD
charge--SET--current' instead of the usual direct SQUID scheme
`magnetic flux--current'. We have analyzed the charge sensitivity
as a function of magnetic flux, gate voltage, Coulomb interaction,
dot asymmetry, and temperature.  The sensitivity of our setup can
be further increased by adding an electromechanical
element~\cite{Knobel}: Applying a large electric field to the
charged nanowire, the change in charge will lead to a mechanical
shift of the wire. This shift can then be detected due to the
change in the capacitance of the compound setup as in
Ref~\cite{Knobel}.  In the present work, we have concentrated on a
single-channel wire in order to demonstrate the effect; the case
of an $n$-channel wire ($n=2$ or $n > 2$) can be analyzed using
the same technique and we plan to study this case in the near
future.

\end{document}